\documentclass[11pt,twoside]{article}
\input{epsf.sty}
\usepackage{mathrsfs}
\usepackage{amsmath}
\usepackage{amssymb}
\usepackage{graphicx}
\usepackage{subfigure}
\newtheorem{proposition}{Proposition}
\newtheorem{lemma}{Lemma}
\newtheorem{corollary}{Corollary}

\title{A synthetical two-component model with peakon solutions}
\author{Baoqiang Xia$^{1,2}$\footnote{E-mail address:
xiabaoqiang@126.com}, Zhijun Qiao$^2$\footnote{ E-mail address:
qiao@utpa.edu}, Ruguang Zhou$^1$\footnote{E-mail address:
zhouruguang@jsnu.edu.cn}
\\ $^{1}$School of Mathematics and Statistics, Jiangsu Normal
University,\\
 Xuzhou, Jiangsu 221116, P. R. China
 \\
$^2$Department of Mathematics, University of Texas-Pan American, \\Edinburg, Texas 78541, USA}

\date{}
\begin{document}
\maketitle
\begin{abstract}

A generalized two-component model with peakon solutions is proposed in this paper. It allows an arbitrary function to be involved in
as well as including some existing integrable peakon equations as special reductions. The generalized two-component system is shown to possess Lax pair and infinitely many conservation laws.
Bi-Hamiltonian structures and peakon interactions are discussed in detail for typical representative equations of the generalized system.
In particular, a new type of $N$-peakon solution, which is not in the traveling wave type, is obtained from the generalized system.

\vspace*{0.2cm}

\end{abstract}
\newpage

\section{Introduction}

In recent years, the Camassa-Holm (CH) equation \cite{CH}
\begin{eqnarray}
m_t=2m u_x+m_xu, \quad m=u-u_{xx}+k,
\label{CH}
\end{eqnarray}
($k$ is an arbitrary constant) derived by Camassa and Holm \cite{CH} as a shallow water
wave model, has attracted much attention and various studies.
The CH equation admits Lax representation \cite{CH}, bi-Hamiltonian structure \cite{FF1,OR}, and is integrable by the inverse scattering transformation \cite{C3}.
Also it possesses multiple peaked soliton solutions \cite{CH,BSS} and algebro-geometric solutions \cite{GH,Q3}.
The most interesting feature of the CH equation is that it admits peaked
soliton (peakon) solutions in the case of $k=0$ \cite{CH,BSS}. A peakon is a weak
solution in some Sobolev space with corner at its crest.
The stability and analysis study of peakons were discussed in several references \cite{CS2}-\cite{JR}.

The interesting characteristics of the CH equation stimulated more people to search new integrable models which admit peakon solutions.
 Among them, for example, there are:

1.  the Degasperis-Procesi (DP) equation \cite{DP1}
\begin{eqnarray}
m_t=3m u_x+m_xu, \quad m=u-u_{xx},
\label{DP}
\end{eqnarray}
which was shown integrable with Lax pair and bi-Hamiltonian structure in \cite{DP2}, multi-peakon dynamics in \cite{DP3}, and shocks formation in \cite{DP4};

2. the cubic nonlinear peakon equation - the FORQ equation \cite{Fo,OR,Fu}
\begin{eqnarray}
 m_t=\left[ m(u^2-u^2_x)\right]_x, \quad  m=u-u_{xx},\label{mCH}
\end{eqnarray}
which was shown to have Lax pair and bi-Hamiltonian structure in \cite{Q1}, and peakon solutions in \cite{GLOQ};

3. the Novikov's cubic nonlinear equation \cite{NV1}
\begin{eqnarray}
 m_t=u^2m_x+3uu_xm, \quad  m=u-u_{xx},\label{cCHN}
\end{eqnarray}
which was shown integrable with Lax pair, bi-Hamiltonian structure and conservation laws in \cite{HW1}; and

4. the generalized CH equation with both quadratic and cubic nonlinearity \cite{Fo,Fu,QXL}
\begin{eqnarray}
 m_t=\frac{1}{2}k_1\left[ m(u^2-u^2_x)\right]_x+\frac{1}{2}k_2(2 m u_x+ m_xu), \quad  m=u-u_{xx},\label{gCH}
\end{eqnarray}
where $k_1$ and $k_2$ are two arbitrary constants. Equation (\ref{gCH}) was proven to possess Lax pair, conservation laws and peakon solutions in \cite{QXL}.

Equation (\ref{gCH}) is actually a linear combination of CH equation (\ref{CH}) and cubic nonlinear equation (\ref{mCH}). This structure is very similar to the Gardner equation,
known as a linear combination of KdV and mKdV equations. Thus equation (\ref{gCH}) is the dual system of the Gardner equation from the viewpoint of tri-Hamiltonian duality \cite{OR,Fu}.
In the literature \cite{Fo}, a more generalized version of equation (\ref{gCH}) was derived by Fokas from the two-dimensional hydrodynamical equations for surface waves.
We also notice that by some appropriate rescaling or gauge transformations equation (\ref{gCH}) is equivalent to equation (\ref{mCH}).

All equations shown above are scalar integrable peakon models. Another important task is to find  integrable multi-component peakon systems to enrich the theory of soliton and integrable systems.
For example, the integrable two-component CH equations are proposed in \cite{OR, CLZ, Fa,HI}. The integrable two-component forms of the cubic peakon systems (\ref{mCH}) and (\ref{cCHN}) are presented in \cite{SQQ,XQ,GX2}.

In addition to the integrable generalizations of peakon equations, there are also some works appearing to study the non-integrable generalizations of peakon equations. The most well-known example is the
so-called $b$-family equation by Holm and Staley \cite{be1,be2}
\begin{equation}
m_t=b mu_x+m_xu, \quad m=u-u_{xx},
\label{bCH}
\end{equation}
where $b$ is an arbitrary constant. The case of $b=2$ is exactly the CH equation, while the case of $b=3$ recovers the DP equation. According to various tests for integrability,
it is known that the cases of $b = 2$ and $b=3$ are the only integrable equations within this family \cite{MN}-\cite{H4}.
But for any $b$, all those equations admit peakon solutions \cite{be2}. Holm and Staley also studied peakon dynamics
of (\ref{bCH}) for different values of $b$ and discussed how they behave with changing $b$ \cite{be2}.
In the literature \cite{P1}, Popowicz proposed
a two-component system, which can be considered as a coupling between the CH equation
and the DP equation. 
Later, Hone and Irle showed that the  two-component Popowicz system is non-integrable, but admits single-peakon solution as well as multi-peakon solutions \cite{H5}.

In this paper, we propose the following generalized version of the two-component peakon system
\begin{eqnarray}
\left\{\begin{array}{l}
m_t=(mH)_x+mH-\frac{1}{2}m(u-u_x)(v+v_x),
\\
n_t=(nH)_x-nH+\frac{1}{2}n(u-u_x)(v+v_x),
\\
m=u-u_{xx},
\\
n=v-v_{xx},
\end{array}\right.
\label{geq}
\end{eqnarray}
where $H$ is an arbitrary function of $u$, $v$ and their derivatives.
As $v=2$ and $H=u$, equation (\ref{geq}) is reduced to the CH equation (\ref{CH}).
As $v=2u$ and $H=(u^2-u^2_x)$, equation (\ref{geq}) is reduced to the FORQ equation (\ref{mCH}).
As $v=k_1u+k_2$ and $H=\frac{1}{2}[k_1(u^2-u^2_x)+k_2u]$, equation (\ref{geq}) is cast into the generalized CH equation (\ref{gCH}).
Thus, equation (\ref{geq}) is a kind of the two-component generalization of equations (\ref{CH}), (\ref{mCH}) and (\ref{gCH}). We show that the generalized system (\ref{geq}) possesses an $sl(2)$-valued Lax pair and infinitely many conservation laws. Since the arbitrary function $H$ is involved in (\ref{geq}), we do not expect all those equations have bi-Hamiltonian structures in general.
Nevertheless, we demonstrate that for some special choices of $H$ we may find the corresponding bi-Hamiltonian structures.
Such a system is interesting, because we may obtain quite a large number of integrable peakon equations by choosing different $H$.
We take some examples to discuss in detail the bi-Hamiltonian structures and the peakon interactions for some equations in the family (\ref{geq}).
From the equations in the family (\ref{geq}), we obtain a new type of $N$-peakon solution which is not presented in the traveling wave type.

The whole paper is organized as follows. Section 2 provides the Lax pair and conservation laws for the system (\ref{geq}). Section 3 studies the bi-Hamiltonian structures and the multi-peakon solutions of some two-component equations in the family (\ref{geq}). Section 4 supplies a proof for the bi-Hamiltonian property in each example discussed in section 3. Some conclusions and open problems are addressed in section 5.

\section{ Lax pair and conservation laws}

Let us consider a pair of $2\times 2$ matrix spectral problems of the following type
\begin{eqnarray}
\phi_x&=&
U\phi,
~~ U=\frac{1}{2}\left( \begin{array}{cc} -1 & \lambda m\\
 -\lambda n &  1 \\ \end{array} \right),
 \label{lps}
 \\
\phi_t&=&
V\phi,
~~ V=-\frac{1}{2}\left( \begin{array}{cc} \lambda^{-2}+\frac{1}{2}(u-u_x)(v+v_x) & -\lambda^{-1}(u-u_x)-\lambda mH \\
 \lambda^{-1}(v+v_x)+\lambda nH & -\lambda^{-2}-\frac{1}{2}(u-u_x)(v+v_x)  \\ \end{array} \right),
 \label{lpt}
\end{eqnarray}
where
\begin{eqnarray}
\begin{split}
\phi=(\phi_{1},~\phi_{2})^T, \quad m=u-u_{xx}, \quad n=v-v_{xx},
\end{split}
\end{eqnarray}
$\lambda$ is a spectral parameter and $H$ is an arbitrary function of $u$, $v$ and their derivatives.

It is easy to see that the compatibility condition of (\ref{lps}) and (\ref{lpt}) reads
\begin{eqnarray}
U_t-V_x+[U,V]=0.
\label{cc}
\end{eqnarray}
Substituting the expressions of (\ref{lps}) and (\ref{lpt}) into (\ref{cc}), we immediately find that (\ref{cc}) is nothing
but equation (\ref{geq}). Thus, (\ref{lps}) and (\ref{lpt}) compose of a Lax pair of equation (\ref{geq}).

\vspace*{0.2cm}
{\bf Remark 1.}
Our generalized system with an arbitrary function $H$ involved does admit an $sl(2)$-valued Lax representation.
System (\ref{geq}) is produced by the compatibility condition (\ref{cc}) of the spectral problems (\ref{lps}) and (\ref{lpt}) where such an arbitrary function is included in $V$ part. The arbitrary function $H$ is able to appear 
because the Lax equation (\ref{cc}) is an overdetermined system by choosing the appropriate $V$ (dependent on $\lambda$) to match $U$.
\vspace*{0.2cm}

Next, let us construct the conservation laws for system (\ref{geq}) by using spectral problems (\ref{lps}) and (\ref{lpt}).
Let $\omega=\frac{\phi_2}{\phi_1}$, then it follows from (\ref{lps}) that $\omega$ satisfies the Riccati equation
\begin{eqnarray}
\omega_x=-\frac{1}{2}\lambda m \omega^2+ \omega-\frac{1}{2}\lambda n.
\label{ric}
\end{eqnarray}
Based on (\ref{lps}) and (\ref{lpt}), we obtain
\begin{eqnarray}
\begin{split}
(\ln \phi_1)_x&=-\frac{1}{2}+\frac{1}{2}\lambda m\omega,
\\
\quad (\ln \phi_1)_t&=-\frac{1}{2}\left[\lambda^{-2} -\lambda^{-1}(u-u_x)\omega+\frac{1}{2}(u-u_x)(v+v_x)-\lambda mH\omega\right],
\label{lnp}
\end{split}
\end{eqnarray}
which generates the following conservation law of equation (\ref{geq})
\begin{eqnarray}
\rho_t=A_x,
\label{CL}
\end{eqnarray}
where
\begin{eqnarray}
\begin{split}
\rho&=m\omega,
\\A&=-\frac{1}{2}\lambda^{-1}(u-u_x)(v+v_x)+\lambda^{-2}(u-u_x)\omega+mH\omega.
\end{split}
\label{rj}
\end{eqnarray}
Usually $\rho$ and $A$ are called a conserved density and an associated flux,
respectively.

We are able to derive the explicit forms of conservation densities by expanding $\omega$ in powers of $\lambda$ in two ways.
The first one is to expand $\omega$ in terms of negative powers of $\lambda$ as
\begin{equation}
\omega=\sum_{j=0}^{\infty}\omega_j\lambda^{-j}.\label{oe1}
\end{equation}
By substituting (\ref{oe1}) into (\ref{ric}) and equating the coefficients of powers of $\lambda$, we arrive at
\begin{eqnarray}
\begin{split}
\omega_{0}&=\sqrt{-\frac{n}{m}}, \qquad \omega_{1}=\frac{mn_x-m_xn-2mn}{2m^2n},
\\
\omega_{j+1}&=\frac{1}{m\omega_0}\left[\omega_j-\omega_{j,x}-\frac{1}{2}m\sum_{i+k=j+1,~1\leq i,k\leq j}\omega_i\omega_k\right],\quad j\geq 1.
\end{split}
\label{wj}
\end{eqnarray}
Inserting (\ref{oe1}) and (\ref{wj}) into (\ref{rj}), we obtain the following infinitely many conserved densities and the associated fluxes
\begin{eqnarray}
\begin{split}
\rho_{0}&=\sqrt{-mn}, ~~~~ A_0=H\sqrt{-mn},
\\
\rho_{1}&=\frac{mn_x-m_xn-2mn}{2mn}, ~~~~ A_1=-\frac{1}{2}(u-u_x)(v+v_x)+\frac{(mn_x-m_xn-2mn)H}{2mn},
\\
\rho_{j}&=m\omega_j, ~~~~A_{j}=(u-u_x)\omega_{j-2}+mH\omega_j,\quad j\geq 2,
\end{split}
\label{rjj}
\end{eqnarray}
where $\omega_j$ is given by (\ref{wj}).

The second expansion of $\omega$ is in the positive powers of $\lambda$ as
\begin{equation}
\omega=\sum_{j=0}^{\infty}\omega_j\lambda^{j}.\label{oe2}
\end{equation}
Substituting (\ref{oe2}) into (\ref{ric}) and comparing powers of $\lambda$ lead to
\begin{eqnarray}
\omega_{2j}&=&0, \quad j\geq 0,
\label{wj21}
\\
\omega_{1}&=&\frac{1}{2}(v+v_x) , \quad \omega_{2j+1}-\omega_{2j+1,x}=\frac{1}{2}m\sum_{i+k=2j,~0\leq i,k\leq 2j}\omega_i\omega_k,\quad j\geq 1.
\label{wj22}
\end{eqnarray}
From formula (\ref{wj21}), we know
\begin{eqnarray}
\rho_{2j}=0, \quad A_{2j}=0, \quad j\geq 0,
\label{rjj2}
\end{eqnarray}
which means the even-index conserved densities and associated fluxes are trivial.
From formula (\ref{wj22}), we arrive at the odd-index conserved densities and associated fluxes
\begin{eqnarray}
\begin{split}
\rho_{1}&=\frac{1}{2}m(v+v_x), ~~~~ A_1=(u-u_x)\omega_3+\frac{1}{2}m(v+v_x)H,
\\
\rho_{2j+1}&=m\omega_{2j+1}, ~~~~A_{2j+1}=(u-u_x)\omega_{2j+3}+mH\omega_{2j+1},\quad j\geq 1,
\end{split}
\label{rjj3}
\end{eqnarray}
where the odd-index $\omega_{2j+1}$ is defined by the recursion relation
\begin{eqnarray}
\omega_{2j+1}=\frac{1}{2}(1-\partial_x)^{-1}\left(m\sum_{i+k=2j,~0\leq i,k\leq 2j}\omega_i\omega_k\right),\quad j\geq 1.
\label{wj23}
\end{eqnarray}
We should remark that the relation (\ref{wj23}) shows the nontrivial high-order conserved densities in the sequence (\ref{rjj3}) may involve in nonlocal expressions in $u$ and $v$.
However, the conserved densities in the sequence (\ref{rjj}) are local ones.

\vspace*{0.2cm}
{\bf Remark 2.}
The expressions (\ref{rjj}) and (\ref{rjj3}) show that all members in our generalized system possess the same conserved quantities but different conserved fluxes.
This is because the conserved quantities are derived from the Riccati equation (\ref{ric}) that only depends on the spatial part of the Lax representation which keeps the same for all members in the family;
while the conserved fluxes rely on the temporal part of the Lax representation which changes for different members.

\section{ Two-component peakon systems}
The two-component system (\ref{geq}) is of great interest because different choices of $H$ lead to different peakon equations.
Let us discuss some special cases in the following examples.

{\bf Example 1. A new integrable system with a new type of peakon solutions}

Taking $H=0$ in equation (\ref{geq}) gives rise to the following integrable two-component model
\begin{eqnarray}
\left\{\begin{array}{l}
m_t=-\frac{1}{2}m(u-u_x)(v+v_x),
\\
n_t=\frac{1}{2}n(u-u_x)(v+v_x),
\\
m=u-u_{xx}, ~~n=v-v_{xx}.
\end{array}\right.
\label{teqh0}
\end{eqnarray}
This model can be rewritten as the following bi-Hamiltonian form
\begin{eqnarray}
\left(m_t,~n_t\right)^{T}=J \left(\frac{\delta H_2}{\delta m},~\frac{\delta H_2}{\delta n}\right)^{T}=K \left(\frac{\delta H_1}{\delta m},~\frac{\delta H_1}{\delta n}\right)^{T},\label{BH21}
\end{eqnarray}
where
\begin{eqnarray}
J&=&\left( \begin{array}{cc} 0 & -\partial-1 \\
 -\partial+1  & 0 \\ \end{array} \right),
~~K=\left( \begin{array}{cc}  -m\partial^{-1}m &  m\partial^{-1}n \\
n\partial^{-1}m  & -n\partial^{-1}n \\ \end{array} \right),
\label{JK0}
\\
H_1&=&\frac{1}{2}\int_{-\infty}^{+\infty}m(v+v_x)dx,~~ H_2=\frac{1}{4}\int_{-\infty}^{+\infty}(u-u_{x})^2(v+v_x)ndx.
\label{J0}
\end{eqnarray}
In section 4, we will provide a detailed proof for the compatibility of the Hamiltonian pairs $J$ and $K$ in this example and the next three examples.

\begin{figure}
\begin{minipage}[t]{0.5\linewidth}
\centering
\includegraphics[width=2.2in]{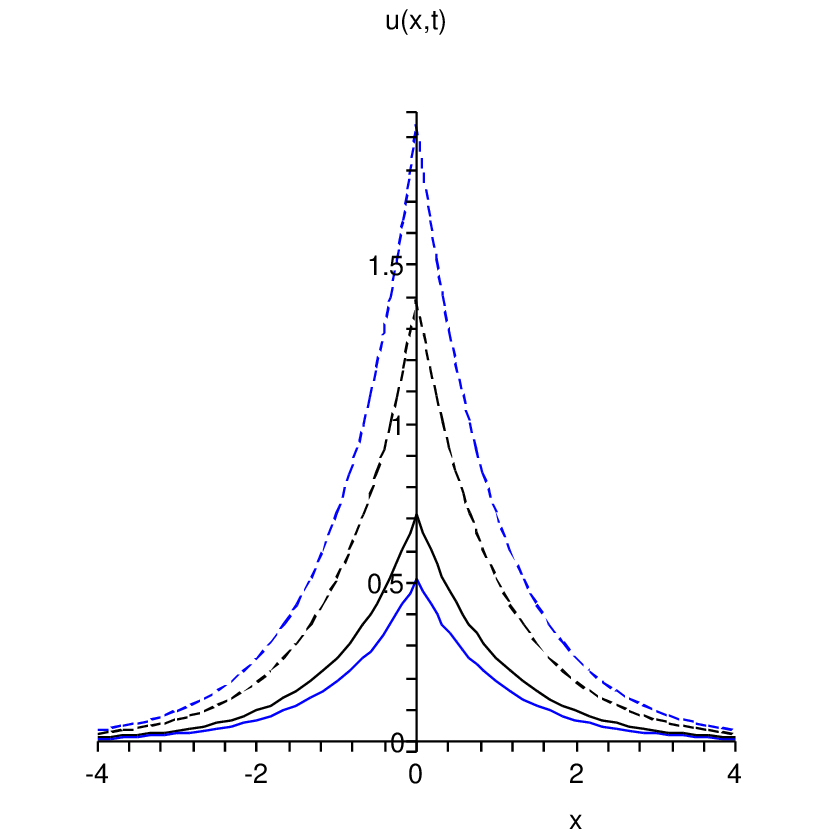}
\caption{\small{The single-peakon solution given by (\ref{steqh0}) with $A_1=A_2=1$ and $A_3=0$. Solid line: $u(x,t)$; Dashed line: $v(x,t)$; Black: $t=1$; Blue: $t=2$.}}
\label{F11}
\end{minipage}
\hspace{2.0ex}
\begin{minipage}[t]{0.5\linewidth}
\centering
\includegraphics[width=2.2in]{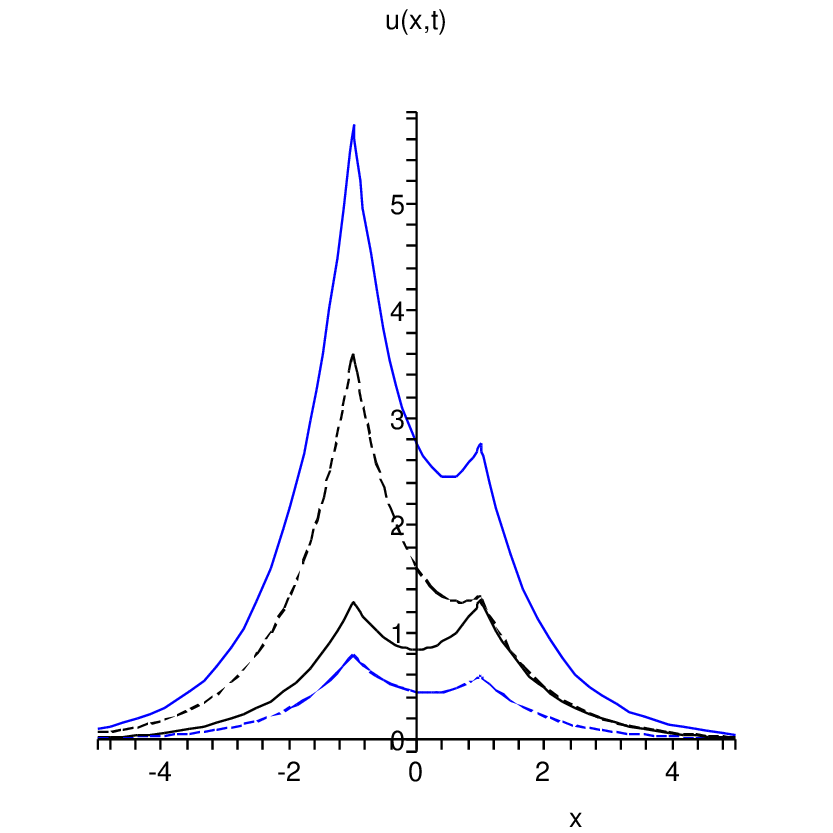}
\caption{\small{The two-peakon solution given by (\ref{41uv2}). Solid line: $u(x,t)$; Dashed line: $v(x,t)$; Black: $t=0$; Blue: $t=-1$.}}
\label{F12}
\end{minipage}
\end{figure}

Let us assume that (\ref{teqh0}) has the following one-peakon solution
\begin{eqnarray}
u=p_1(t)e^{-\mid x-q_1(t)\mid},\quad v=r_1(t)e^{-\mid x-q_1(t)\mid}, \label{teqh0OP}
\end{eqnarray}
where $p_1(t)$, $r_1(t)$ and $q_1(t)$ are functions of $t$ needed to be determined.
Substituting (\ref{teqh0OP}) into (\ref{teqh0}) and integrating against the test function with support around the peak, we obtain
\begin{eqnarray}
p_{1,t}=-\frac{1}{3}p_1^2r_1,\qquad
r_{1,t}=\frac{1}{3}p_1r_1^2, \qquad
q_{1,t}=0,
\label{opteqh0}
\end{eqnarray}
which yields
\begin{eqnarray}
p_{1}(t)=A_2e^{-\frac{1}{3}A_1t},\qquad
r_{1}(t)=\frac{A_1}{A_2}e^{\frac{1}{3}A_1t}, \qquad
q_{1}(t)=A_3,
\label{sopteqh0}
\end{eqnarray}
where $A_1$, $A_2$, and $A_3$ are integration constants.
Thus, we obtain the peakon solutions as follows
\begin{eqnarray}
u(x,t)=A_2e^{-\frac{1}{3}A_1t}e^{-|x-A_3|},\qquad
v(x,t)=\frac{A_1}{A_2}e^{\frac{1}{3}A_1t}e^{-|x-A_3|}.
\label{steqh0}
\end{eqnarray}
This pair of single-peakon solutions is not presented in the traveling wave type, because the peakon position $ q_1(t)=A_3$ is stationary. To the best of our knowledge, almost all integrable peakon models have single peakons  which are of traveling wave type. So, we find a new integrable peakon system (\ref{teqh0}) whose peakon solution is not in traveling wave type. See Figure \ref{F11} for the profile of the new single-peakon solution.
We remark that the amplitudes of the peakons of equation (\ref{teqh0}) grow/decay exponentially with time. Recently, Lundmark and Szmigielski \cite{LS} found that the Geng-Xue two-component system \cite{GX2} has a
similar type of peakons (with amplitudes exponentially growing/decaying with time).

Let us suppose the $N$-peakon solution in the form of
\begin{eqnarray}
u(x,t)=\sum_{j=1}^N p_j(t)e^{-\mid x-q_j(t)\mid}, ~~v(x,t)=\sum_{j=1}^N r_j(t)e^{-\mid x-q_j(t)\mid}.
\label{NP}
\end{eqnarray}
By substituting (\ref{NP}) into (\ref{teqh0}) and integrating against test functions, we obtain the $N$-peakon dynamic system of (\ref{teqh0})
\begin{eqnarray}
\left\{\begin{split}
q_{j,t}=&0,\\
p_{j,t}=&\frac{1}{6}p_j^2r_j+\frac{1}{2}p_j\sum_{i,k=1}^Np_ir_k\left(sgn(q_j-q_i)+1\right)\left(sgn(q_j-q_k)-1\right)e^{ -\mid q_j-q_i\mid-\mid q_j-q_k\mid},\\
r_{j,t}=&-\frac{1}{6}p_jr_j^2-\frac{1}{2}r_j\sum_{i,k=1}^Np_ir_k\left(sgn(q_j-q_i)+1\right)\left(sgn(q_j-q_k)-1\right)e^{ -\mid q_j-q_i\mid-\mid q_j-q_k\mid}.
\end{split}\right.
\label{teqh0NP}
\end{eqnarray}
In the above formula, $q_{j,t}=0$ implies that the peak position does not change along with the time $t$.

For $N=2$, solving (\ref{teqh0NP}) leads to
\begin{eqnarray}
\left\{\begin{array}{l}
q_1(t)=A_4, ~~~~q_2(t)=A_5,\\
p_{1}(t)=A_6e^{-\frac{1}{3}A_1t-\frac{e^{-|A_4-A_5|}}{2}\left[\frac{3A_1(1+sgn(A_4-A_5))}{(A_1-A_2)A_3}e^{\frac{1}{3}(A_1-A_2)t}
-\frac{3A_2A_3(1-sgn(A_4-A_5))}{A_1-A_2}e^{-\frac{1}{3}(A_1-A_2)t}\right]},\\
p_{2}(t)=\frac{p_1}{A_3}e^{\frac{1}{3}(A_1-A_2)t},\\
r_{1}(t)=\frac{A_1}{p_1}, ~~~~r_{2}(t)=\frac{A_2}{p_2}, \\
\end{array}\right. \label{stpteqh0}
\end{eqnarray}
where $A_1$, $A_2$, $\cdots$, $A_6$ are integration constants. If $A_4=A_5$, it is reduced to the one-peakon solution. If $A_4\neq A_5$, this two-peakon  solution will never collide because $q_1(t)\neq q_2(t)$ for any $t$.
In particular, for $A_1=A_3=A_4=-A_5=A_6=1$ and $A_2=4$, the two-peakon becomes
\begin{eqnarray}
\left\{\begin{array}{l}
u(x,t)=e^{-\frac{1}{3}t+e^{-t-2}}e^{-\mid x-1\mid}+e^{-\frac{4}{3}t+e^{-t-2}}e^{-\mid x+1\mid},\\
v(x,t)=e^{\frac{1}{3}t-e^{-t-2}}e^{-\mid x-1\mid}+4e^{\frac{4}{3}t-e^{-t-2}}e^{-\mid x+1\mid}.\\
\end{array}\right. \label{41uv2}
\end{eqnarray}
See Figure \ref{F12} for the profile of the two-peakon dynamics for the potentials $u(x,t)$ and $v(x,t)$.

\vspace*{0.2cm}
{\bf Remark 3.}
We point out that from (\ref{teqh0}) one may conclude $(mn)_t =0$ and
thus $mn = f(x)$, where $f(x)$ is a free function of $x$. It then follows that $v=(1-\partial^2)^{-1}(\frac{f(x)}{m})$. This means we can remove the component $v$ in equation (\ref{teqh0})
and thus write (\ref{teqh0}) in the form of a single field equation. However, the resulting single field equation involves in nonlocal expressions and a free function $f(x)$.
Guided by this, we still write the equation associated with the case of $H=0$ in the form of (\ref{teqh0}).
\vspace*{0.2cm}

{\bf Example 2. The integrable two-component system proposed in \cite{XQ}}

By choosing $H=\frac{1}{2}\left(uv-u_xv_x\right)$, we obtain
\begin{eqnarray}
\left\{\begin{array}{l}
m_t=\frac{1}{2}[m\left(uv-u_xv_x\right)]_x-\frac{1}{2}m\left(uv_x-u_xv\right),
\\
n_t=\frac{1}{2}[n\left(uv-u_xv_x\right)]_x+\frac{1}{2}n\left(uv_x-u_xv\right),
\\
m=u-u_{xx}, ~~n=v-v_{xx},
\end{array}\right.
\label{teqhuv1}
\end{eqnarray}
which is exactly the dispersionless version of the system we derived in \cite{XQ}. This system possesses the bi-Hamiltonian form
\begin{eqnarray}
\left(m_t,~ n_t\right)^{T}=J \left(\frac{\delta H_2}{\delta m},~\frac{\delta H_2}{\delta  n}\right)^{T}=K \left(\frac{\delta H_1}{\delta m},~\frac{\delta H_1}{\delta  n}\right)^{T},\label{BH22}
\end{eqnarray}
where
\begin{eqnarray}
J&=&\left( \begin{array}{cc} 0 &  \partial^2-1 \\
1-\partial^2  & 0 \\ \end{array} \right), ~~K=\left( \begin{array}{cc} \partial m\partial^{-1}m\partial-m\partial^{-1}m &  \partial m\partial^{-1} n\partial+m\partial^{-1} n \\
\partial n\partial^{-1} m\partial+ n\partial^{-1} m & \partial n\partial^{-1} n\partial- n\partial^{-1} n \\ \end{array} \right),
\label{JK1}
\\
H_1&=&\frac{1}{2}\int_{-\infty}^{+\infty}(uv+u_xv_x)dx, ~~H_2=\frac{1}{4}\int_{-\infty}^{+\infty}(u^2v_x+u_x^2v_x-2uu_xv)n dx.
\label{H}
\end{eqnarray}
In \cite{XQ}, we have derived the one-peakon of (\ref{teqhuv1})
\begin{eqnarray}
u(x,t)=c_1e^{-|x+\frac{1}{3}c_1c_2t|}, \qquad v(x,t)=c_2e^{-|x+\frac{1}{3}c_1c_2t|},
\label{opteqhuv1}
\end{eqnarray}
where $c_1$ and $c_2$ are two arbitrary integration constants.
We also investigated the $N$-peakon dynamical system. In particular, the two-peakon solution was given explicitly and the collisions are discussed (for details, see \cite{XQ}).

{\bf Example 3. A new integrable two-component peakon system with the same bi-Hamiltonian operators as (\ref{teqhuv1}) but different Hamiltonian functions}

Taking $H=\frac{1}{2}\left(uv_x-u_xv\right)$, we arrive at
\begin{eqnarray}
\left\{\begin{array}{l}
m_t=\frac{1}{2}\left[m(uv_x-u_xv)\right]_x-\frac{1}{2}m\left(uv-u_xv_x\right),
\\
n_t=\frac{1}{2}\left[n(uv_x-u_xv)\right]_x+\frac{1}{2}n\left(uv-u_xv_x\right),
\\
m=u-u_{xx}, ~~n=v-v_{xx}.
\end{array}\right.
\label{teqhuv2}
\end{eqnarray}
This system can be rewritten as the following bi-Hamiltonian form
\begin{eqnarray}
\left(m_t,~n_t\right)^{T}=J \left(\frac{\delta H_2}{\delta m},~\frac{\delta H_2}{\delta n}\right)^{T}=K \left(\frac{\delta H_1}{\delta m},~\frac{\delta H_1}{\delta n}\right)^{T},\label{BH23}
\end{eqnarray}
where $J$, $K$ are given by (\ref{JK1}), and
\begin{eqnarray}
H_1=\frac{1}{2}\int_{-\infty}^{+\infty}(uv_x+u_xv_{xx})dx, ~~H_2=\frac{1}{4}\int_{-\infty}^{+\infty}(u^2v+u_x^2v-2uu_xv_x)n dx.
\label{H2}
\end{eqnarray}
From (\ref{BH22}) and (\ref{BH23}), we find that equation (\ref{teqhuv1}) and (\ref{teqhuv2}) share the same bi-Hamiltonian operators but with different Hamiltonian functions. We will comment this at the end of this
example (see Remark 4 below).

\begin{figure}
\begin{minipage}[t]{0.3\linewidth}
\centering
\includegraphics[width=2.2in]{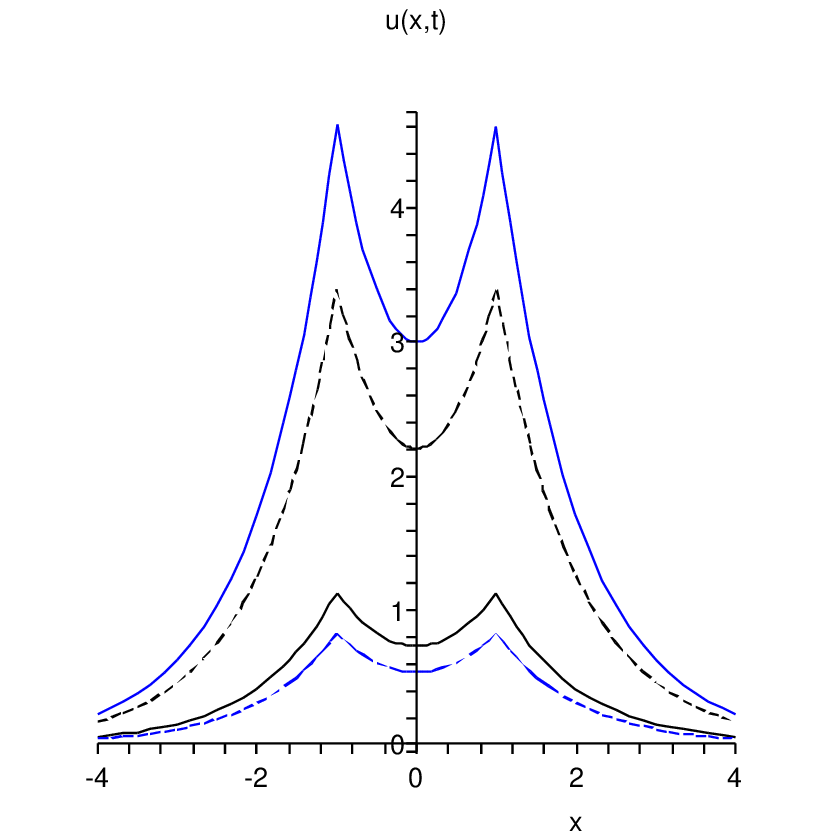}
\caption{\small{The M-shape peakon solution given by (\ref{suv21c1}). Solid line: $u(x,t)$; Dashed line: $v(x,t)$; Black: $t=0$; Blue: $t=-1$.}}
\label{F21}
\end{minipage}
\hspace{1.9ex}
\begin{minipage}[t]{0.3\linewidth}
\centering
\includegraphics[width=2.2in]{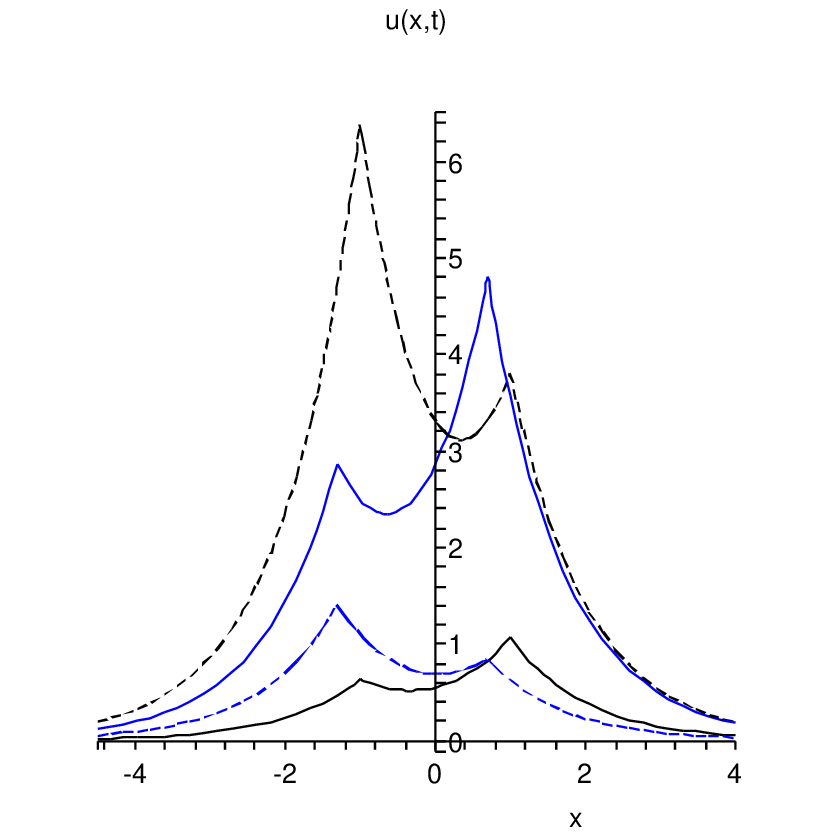}
\caption{\small{The two-peakon solution given by (\ref{suv21c2}). Solid line: $u(x,t)$; Dashed line: $v(x,t)$; Black: $t=0$; Blue: $t=-1$.}}
\label{F22}
\end{minipage}
\hspace{1.9ex}
\begin{minipage}[t]{0.3\linewidth}
\centering
\includegraphics[width=2.2in]{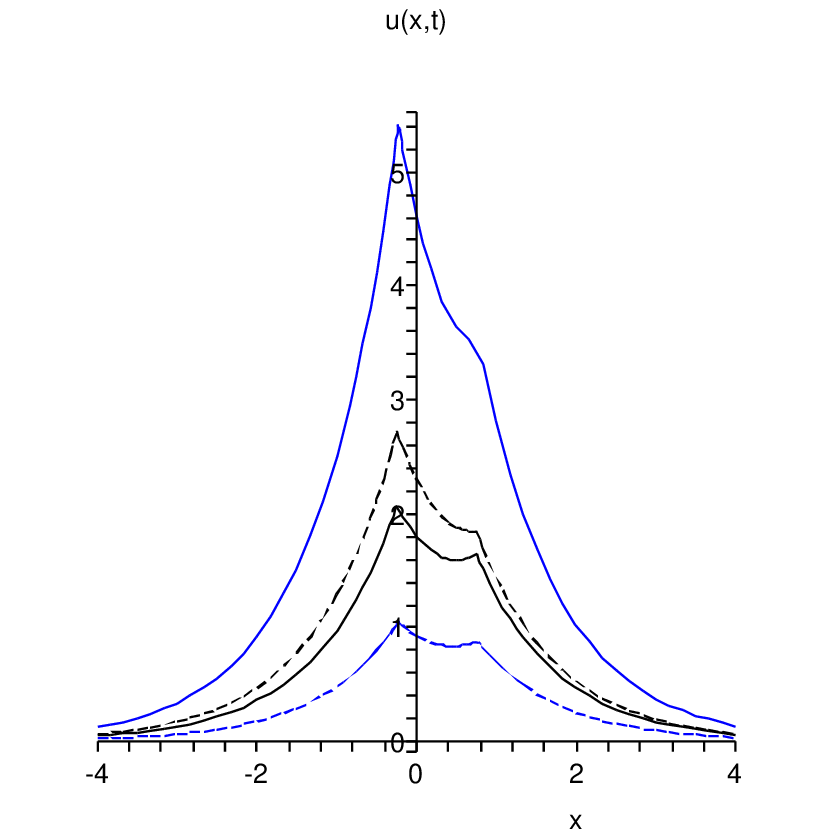}
\caption{\small{The two-peakon solution determined by (\ref{42pq2}). Solid line: $u(x,t)$; Dashed line: $v(x,t)$; Black: $t=-0.5$; Blue: $t=-1$.}}
\label{F23}
\end{minipage}
\end{figure}

Let us study the peakon solutions of this example. By direct calculations, we find that the one-peakon solution of (\ref{teqhuv2}) takes the form as
\begin{eqnarray}
u(x,t)=c_2e^{-\frac{1}{3}c_1t}e^{-|x-c_3|},\qquad
v(x,t)=\frac{c_1}{c_2}e^{\frac{1}{3}c_1t}e^{-|x-c_3|},
\label{e33}
\end{eqnarray}
where $c_1$, $c_2$ and $c_3$ are three integration constants.
In general, we suppose the $N$-peakon solution of (\ref{teqhuv2}) in the form of (\ref{NP}). Then we obtain the $N$-peakon dynamical system of (\ref{teqhuv2})
\begin{eqnarray}
\left\{\begin{split}
p_{j,t}=&\frac{1}{6}p_j^2r_j+\frac{1}{2}p_j\sum_{i,k=1}^Np_ir_k\left(sgn(q_j-q_i)sgn(q_j-q_k)-1\right)e^{ -\mid q_j-q_i\mid-\mid q_j-q_k\mid},\\
r_{j,t}=&-\frac{1}{6}p_jr_j^2-\frac{1}{2}r_j\sum_{i,k=1}^Np_ir_k\left(sgn(q_j-q_i)sgn(q_j-q_k)-1\right)e^{ -\mid q_j-q_i\mid-\mid q_j-q_k\mid},\\
q_{j,t}=&\frac{1}{2}\sum_{i,k=1}^Np_ir_k\left(sgn(q_j-q_k)-sgn(q_j-q_i)\right)e^{ -\mid q_j-q_i\mid-\mid q_j-q_k\mid}.
\end{split}\right.
\label{teqhuv2NP}
\end{eqnarray}
For $N=2$, the two-peakon dynamical system reads as
\begin{eqnarray}
\left\{\begin{array}{l}
p_{1,t}=-\frac{1}{3}p_1^2r_1-\frac{1}{2}p_1\left(p_1r_2+p_2r_1\right)e^{ -\mid q_1-q_2\mid},\\
p_{2,t}=-\frac{1}{3}p_2^2r_2-\frac{1}{2}p_2\left(p_1r_2+p_2r_1\right)e^{ -\mid q_1-q_2\mid},\\
r_{1,t}=\frac{1}{3}p_1r_1^2+\frac{1}{2}r_1\left(p_1r_2+p_2r_1\right)e^{ -\mid q_1-q_2\mid},\\
r_{2,t}=\frac{1}{3}p_2r_2^2+\frac{1}{2}r_2\left(p_1r_2+p_2r_1\right)e^{ -\mid q_1-q_2\mid},\\
q_{1,t}=\frac{1}{2}\left(p_1r_2-p_2r_1\right)sgn(q_1-q_2)e^{ -\mid q_1-q_2\mid},\\
q_{2,t}=q_{1,t}.\\
\end{array}\right. \label{tpteqhuv2}
\end{eqnarray}
From the first four equations of (\ref{tpteqhuv2}), we may conclude $p_{1}(t)r_{1}(t)=A_1$ and $p_{2}(t)r_{2}(t)=A_2$  where $A_1$ and $A_2$ are two integration constants. From the last two equations of (\ref{tpteqhuv2}), we know $q_{2}(t)=q_{1}(t)-B_1$ where $B_1$ is a nonzero constant, which indicates that the two-peakon will never collide.
For $A_1=A_2$, we have
\begin{eqnarray}
\left\{\begin{array}{l}
p_{1}(t)=De^{\left[-\frac{1}{3}A_1-\frac{1}{2}\left(A_1C_1+\frac{A_1}{C_1}\right)e^{-\mid B_1\mid}\right]t},\\
p_{2}(t)=\frac{p_1(t)}{C_1},~~~~r_{1}(t)=\frac{A_1}{p_1(t)}, ~~~~r_{2}=\frac{A_1C_1}{p_1(t)}, \\
q_{1}(t)=\frac{1}{2}\left[(A_1C_1-\frac{A_1}{C_1})sgn(B_1)e^{-\mid B_1\mid}\right]t+\frac{B_1}{2},\\
q_{2}(t)=q_1(t)-B_1,\\
\end{array}\right. \label{stpteqhuv21}
\end{eqnarray}
where $B_1$, $C_1$, and $D$ are three integration constants. For example, choosing $C_1=D=1$, $B_1=2$, $A_1=3$, we have
$p_{2}(t)=p_{1}(t)=e^{-(3e^{-2}+1)t},~~r_{2}(t)=r_{1}(t)=3e^{(3e^{-2}+1)t}$. Thus, the two-peakon solution accordingly reads as
\begin{eqnarray}
\left\{\begin{array}{l}
u(x,t)=e^{-(3e^{-2}+1)t}\left(e^{-\mid x-1\mid}+e^{-\mid x+1\mid}\right),
\\
v(x,t)=3e^{(3e^{-2}+1)t}\left(e^{-\mid x-1\mid}+e^{-\mid x+1\mid}\right),
\end{array}
\right.
\label{suv21c1}
\end{eqnarray}
which are apparently M-shape peakon solutions with two peaks (see Figure \ref{F21} for details).
If choosing $C_1=B_1=2$, $D=1$, $A_1=3$, then we have the following two-peakon solution
\begin{eqnarray}
\left\{
\begin{array}{l}
u(x,t)=\frac{1}{2}e^{-(\frac{15}{4}e^{-2}+1)t}\left(2e^{-\mid x-\frac{9}{4}e^{-2}t-1\mid}+e^{-\mid x-\frac{9}{4}e^{-2}t+1\mid}\right),
\\
v(x,t)=3e^{(\frac{15}{4}e^{-2}+1)t}\left(e^{-\mid x-\frac{9}{4}e^{-2}t-1\mid}+2e^{-\mid x-\frac{9}{4}e^{-2}t+1\mid}\right).
\end{array}
\right.
\label{suv21c2}
\end{eqnarray}
Figure \ref{F22} shows the profile of this two-peakon solution.

For $A_1\neq A_2$, we obtain the following solution of (\ref{tpteqhuv2})
\begin{eqnarray}
\left\{\begin{array}{l}
p_{1}(t)=B_3e^{-\frac{1}{3}A_1t-\frac{3e^{-|B_1|}}{2(A_1-A_2)}\left(\frac{A_1}{B_2}e^{\frac{1}{3}(A_1-A_2)t}-A_2B_2e^{-\frac{1}{3}(A_1-A_2)t}\right)},\\
p_{2}(t)=\frac{p_1}{B_2}e^{\frac{1}{3}(A_1-A_2)t},\\
r_{1}(t)=\frac{A_1}{p_1}, ~~~~r_{2}=\frac{A_2}{p_2}, \\
q_{1}(t)=-\frac{3sgn(B_1)e^{-|B_1|}}{2(A_1-A_2)}\left[A_2B_2e^{-\frac{1}{3}(A_1-A_2)t}+\frac{A_1}{B_2}e^{\frac{1}{3}(A_1-A_2)t}\right]+B_4,\\
q_{2}(t)=q_1-B_1,\\
\end{array}\right. \label{stpteqhuv22}
\end{eqnarray}
where $A_1$, $A_2$, $B_1$, $B_2$, $B_3$, and $B_4$ are six integration constants. Let us consider a special case of choosing $A_1=B_1=B_2=B_3=1$, $A_2=4$, $B_4=0$. Then we have
\begin{eqnarray}
\left\{\begin{array}{l}
p_{1}=e^{-\frac{1}{3}t+\frac{1}{2}e^{-t-1}-2e^{t-1}},\\
p_{2}=e^{-\frac{4}{3}t+\frac{1}{2}e^{-t-1}-2e^{t-1}},\\
r_{1}=e^{\frac{1}{3}t-\frac{1}{2}e^{-t-1}+2e^{t-1}},\\
r_{2}=4e^{\frac{4}{3}t-\frac{1}{2}e^{-t-1}+2e^{t-1}},\\
q_{1}=\frac{1}{2}e^{-t-1}+2e^{t-1},\\
q_{2}=q_1-1.\\
\end{array}\right. \label{42pq2}
\end{eqnarray}
Figure \ref{F23} shows the dynamics of this two-peakon for the potentials $u(x,t)$ and $v(x,t)$ determined by (\ref{42pq2}).

\vspace*{0.2cm}
{\bf Remark 4.}  It has been shown equations (\ref{teqhuv1}) and (\ref{teqhuv2}) share the same bi-Hamiltonian operators (but with different Hamiltonian functions). In fact, the bi-Hamiltonian operators (\ref{JK1}) generate two hierarchies of equations. To see this, we define Lenard sequence recursively by
\begin{eqnarray*}
Jb_{-k}=Kb_{-k+1}, \quad Jb_0=0, \quad k\in\mathbb{Z^{+}},
\end{eqnarray*}
and the soliton hierarchy by
\begin{eqnarray}
m_{t_{-n}}=Kb_{-n}, \quad n\in\mathbb{Z^{+}}.
\label{hk}
\end{eqnarray}
Let us take an initial value $b_0=(0,0)^T$. Then from $Jb_{-1}=Kb_{0}$,  we may reach
$b_{-1}=\frac{1}{2}(v,u)^T$ or $b_{-1}=\frac{1}{2}(v_x,-u_x)^T$.
For $b_{-1}=\frac{1}{2}(v,u)^T$, the first member $m_{t_{-1}}=Kb_{-1}$ in the hierarchy (\ref{hk}) is just equation (\ref{teqhuv1}).
While for $b_{-1}=\frac{1}{2}(v_x,-u_x)^T$, the first member $m_{t_{-1}}=Kb_{-1}$ is nothing but equation (\ref{teqhuv2}).

\vspace*{0.2cm}

{\bf Remark 5.} Although equations (\ref{teqhuv1}) and (\ref{teqhuv2}) share the same bi-Hamiltonian operators, their peakon dynamics are very different. In the single-peakon case, the peakon solution of (\ref{teqhuv1}) is in the type of traveling wave (see (\ref{opteqhuv1})), while the peakon solution of (\ref{teqhuv2}) is not, since the peak point does not change along with the time $t$ (see (\ref{e33})). In the two-peakon case, the collision of the two-peakon of equation (\ref{teqhuv1}) is discussed in \cite{XQ}, while the two-peakon of equation (\ref{teqhuv2}) never collides since their positions are satisfied with $q_{2}(t)=q_{1}(t)-B_1$, where $B_1$ is a nonzero constant.
\vspace*{0.2cm}

{\bf Example 4. The two-component integrable system proposed by Song, Qu, and Qiao \cite{SQQ}}

Choosing $H=\frac{1}{2}(u-u_x)(v+v_x)$ casts equation (\ref{geq}) into
\begin{eqnarray}
\left\{\begin{array}{l}
m_t=\frac{1}{2}\left[m(u-u_x)(v+v_x)\right]_x,
\\
n_t=\frac{1}{2}\left[n(u-u_x)(v+v_x)\right]_x,
\\
m=u-u_{xx}, ~~n=v-v_{xx},
\end{array}\right.
\label{SQQ}
\end{eqnarray}
which is exactly the equation derived by Song, Qu, and Qiao \cite{SQQ}.
This system possesses a bi-Hamiltonian structure \cite{TL}:
\begin{eqnarray}
\left(m_t,~n_t\right)^{T}=J \left(\frac{\delta H_2}{\delta m},~\frac{\delta H_2}{\delta n}\right)^{T}=K \left(\frac{\delta H_1}{\delta m},~\frac{\delta H_1}{\delta n}\right)^{T},\label{HSSQQ}
\end{eqnarray}
where
\begin{eqnarray}
J&=&\left( \begin{array}{cc} 0 &  \partial^2+\partial \\
-\partial^2+\partial  & 0 \\ \end{array} \right),
~~~~K=\left( \begin{array}{cc} \partial m\partial^{-1}m\partial &  \partial m\partial^{-1} n\partial \\
\partial n\partial^{-1} m\partial & \partial n\partial^{-1} n\partial  \\ \end{array} \right),
\label{JK2}
\\
H_1&=&\frac{1}{2}\int_{-\infty}^{+\infty}m(v+v_x)dx, ~~~~H_2=\frac{1}{4}\int_{-\infty}^{+\infty}(u-u_{x})^2(v+v_x)ndx.
\label{J1}
\end{eqnarray}

In the following, we want to derive the peakon solutions and discuss the peakon interactions for this system. It is easy to check that the one-peakon solution of (\ref{SQQ}) takes the same form as (\ref{opteqhuv1}).
In general, by direct calculations, we can obtain the $N$-peakon dynamical system of (\ref{SQQ}) as follows
\begin{eqnarray}
\left\{\begin{split}
p_{j,t}=&0,\\
r_{j,t}=&0,\\
q_{j,t}=&\frac{1}{6}p_jr_j+\frac{1}{2}\sum_{i,k=1}^Np_ir_k\left(sgn(q_j-q_i)+1\right)\left(sgn(q_j-q_k)-1\right)e^{ -\mid q_j-q_i\mid-\mid q_j-q_k\mid}.
\end{split}\right.
\label{dNcp}
\end{eqnarray}

\begin{figure}
\begin{minipage}[t]{0.5\linewidth}
\centering
\includegraphics[width=2.2in]{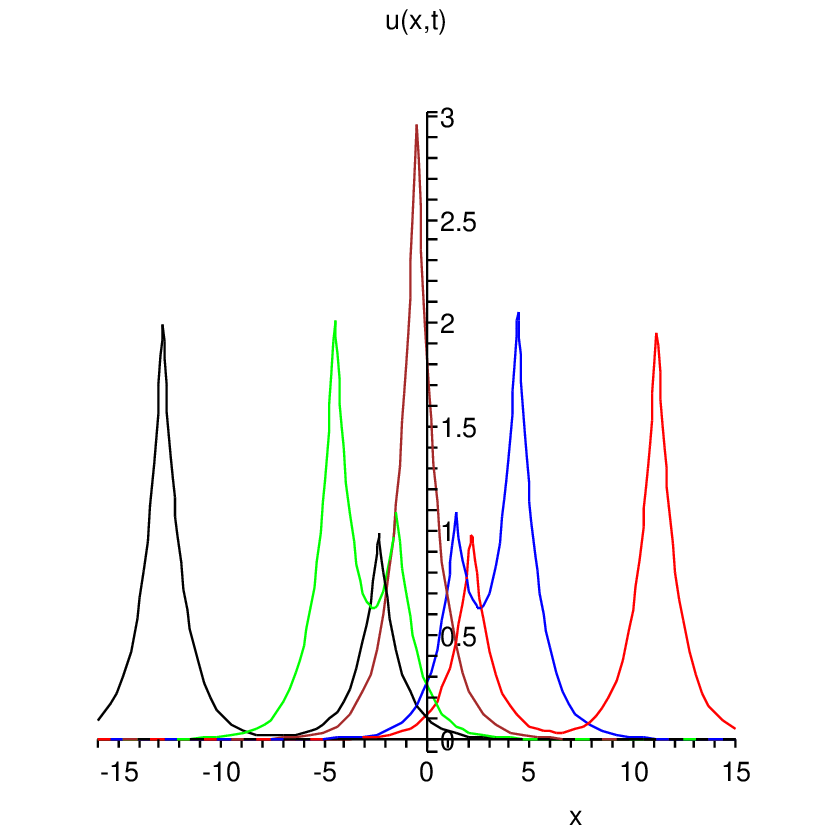}
\caption{\small{ The two-peakon solution for the potential $u(x,t)$ given by (\ref{tpSQQuv}). Red line: $t=-3$; Blue line: $t=-1$; Brown line: $t=0$ (collision); Green line: $t=1$; Black line: $t=3.5$. }}
\label{F34}
\end{minipage}
\hspace{2.0ex}
\begin{minipage}[t]{0.5\linewidth}
\centering
\includegraphics[width=2.2in]{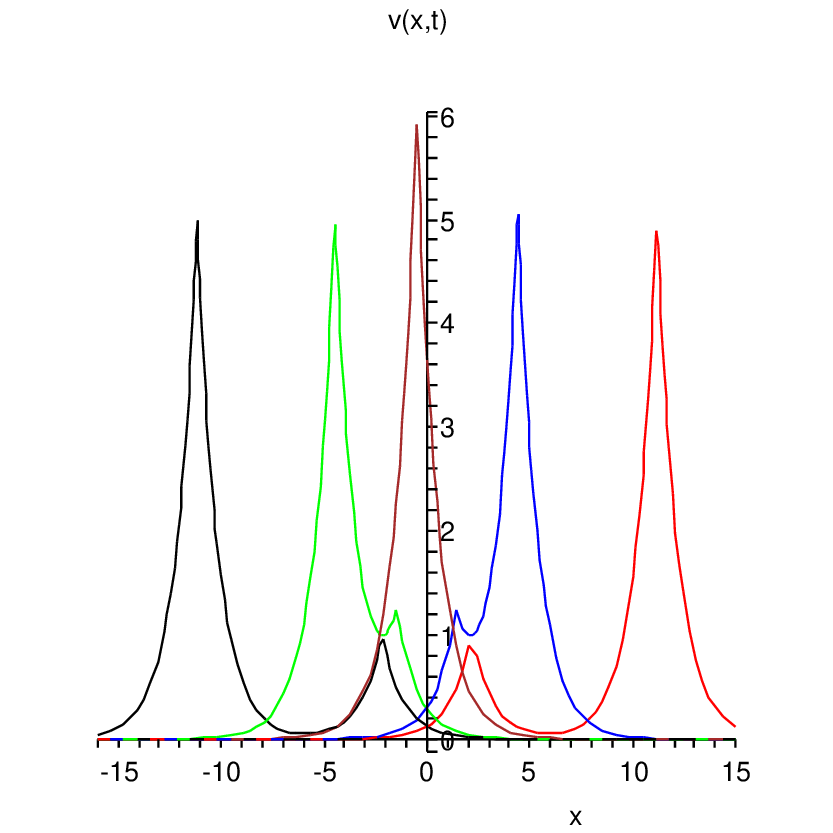}
\caption{\small{ The two-peakon solution for the potential $v(x,t)$ given by (\ref{tpSQQuv}). Red line: $t=-3$; Blue line: $t=-1$; Brown line: $t=0$ (collision); Green line: $t=1$; Black line: $t=3$. }}
\label{F35}
\end{minipage}
\end{figure}

If $N=2$, then the two-peakon system reads as
\begin{eqnarray}
\left\{\begin{array}{l}
p_{1,t}=p_{2,t}=r_{1,t}=r_{2,t}=0,\\
q_{1,t}=-\frac{1}{3}p_1r_1+\frac{1}{2}\left[p_1r_2\left(sgn(q_1-q_2)-1\right)-p_2r_1\left(sgn(q_1-q_2)+1\right)\right]e^{ -\mid q_1-q_2\mid},\\
q_{2,t}=-\frac{1}{3}p_2r_2+\frac{1}{2}\left[p_1r_2\left(sgn(q_1-q_2)-1\right)-p_2r_1\left(sgn(q_1-q_2)+1\right)\right]e^{ -\mid q_1-q_2\mid}.\\
\end{array}\right. \label{tpSQQ}
\end{eqnarray}
From the first equation of (\ref{tpSQQ}), we know
\begin{eqnarray}
p_1=A_1, ~~~~ p_2=A_2, ~~~~r_1=B_1,  ~~~~r_2=B_2,
\label{rtp}
\end{eqnarray}
where $A_1$, $A_2$, $B_1$, and $B_2$ are four integration constants.
If $A_1B_1= A_2B_2$, then we have
\begin{eqnarray}
\left\{\begin{array}{l}
q_{1}(t)=\left\{-\frac{1}{3}A_1B_1+\frac{1}{2}\left[A_1B_2\left(sgn(C_1)-1\right)-A_2B_1\left(sgn(C_1)+1\right)\right]e^{ -\mid C_1\mid}\right\}t+\frac{C_1}{2},\\
q_{2}(t)=q_{1}(t)-C_1.\\
\end{array}\right. \label{tpSQQc1}
\end{eqnarray}
If $A_1B_1\neq A_2B_2$,
then we arrive at
\begin{eqnarray}
\left\{\begin{array}{l}
q_{1}(t)=-\frac{1}{3}A_1B_1t+\Gamma(t),\\
q_{2}(t)=-\frac{1}{3}A_2B_2t++\Gamma(t),
\end{array}\right.
\label{tpSQQc2}
\end{eqnarray}
where
\begin{eqnarray}
\Gamma(t)=\frac{3(A_1B_2+A_2B_1)}{2|A_1B_1-A_2B_2| }sgn(t)\left(e^{- \frac{1}{3}\mid(A_1B_1-A_2B_2)t\mid}-1\right)+\frac{3(A_1B_2-A_2B_1)}{2(A_1B_1-A_2B_2)}e^{- \frac{1}{3}\mid(A_1B_1-A_2B_2)t\mid}.
\label{Gamma}
\end{eqnarray}
In particular, taking $A_1=B_1=1$, $A_2=2$, and $B_2=5$ sends the two-peakon solution to the following form
\begin{eqnarray}
\left\{
\begin{split}
u(x,t)&=e^{-\mid x-q_1(t)\mid}+2e^{-\mid x-q_2(t)\mid},\\
v(x,t)&=e^{-\mid x-q_1(t)\mid}+5e^{-\mid x-q_2(t)\mid},
\end{split}
\right.
\label{tpSQQuv}
\end{eqnarray}
where
\begin{eqnarray}
\left\{\begin{split}
q_{1}(t)&=-\frac{t}{3}+\frac{7}{6}sgn(t)\left(e^{-3|t|}-1\right)-\frac{1}{2}e^{-3|t|},\\
q_{2}(t)&=-\frac{10t}{3}+\frac{7}{6}sgn(t)\left(e^{-3|t|}-1\right)-\frac{1}{2}e^{-3|t|}.
\end{split}
\right.
\label{tpSQQq}
\end{eqnarray}
For the potential $u(x,t)$, the two-peakon collides at the moment $t=0$, since $q_1(0)=q_2(0)=0$.
For $t<0$, the tall and fast peakon with the  amplitude $2$ and peak position $q_2$ chases after
the short and slow peakon with the  amplitude $1$ and peak position $q_1$.
At the moment of $t=0$, the two-peakon overlaps. After the collision ($t>0$), the two-peakon separates,
and the tall and fast peakon surpasses the short and slow one. Similarly,
we may discuss the collision of the two-peakon for the potential $v(x,t)$.
See Figures \ref{F34} and \ref{F35} for the two-peakon dynamics of the potentials $u(x,t)$ and $v(x,t)$.

\section{A proof for the bi-Hamiltonian property}
In this section, we will supply a proof for the bi-Hamiltonian property in each example presented in the above section.
Let us introduce the following basic operators
\begin{eqnarray}
J_1&=&\left( \begin{array}{cc} 0 & -1 \\
 1  & 0 \\ \end{array} \right),
 ~~~~J_2=\left( \begin{array}{cc} 0 & \partial \\
 \partial  & 0 \\ \end{array} \right),
 ~~~~ J_3=\left( \begin{array}{cc} 0 & \partial^2 \\
 -\partial^2  & 0 \\ \end{array} \right),
 \label{Jj}
 \\
K_1&=&\left( \begin{array}{cc}  -m\partial^{-1}m &  m\partial^{-1}n \\
n\partial^{-1}m  & -n\partial^{-1}n \\ \end{array} \right),
~~~~K_2=\left( \begin{array}{cc} \partial m\partial^{-1}m\partial &  \partial m\partial^{-1} n\partial \\
\partial n\partial^{-1} m\partial & \partial n\partial^{-1} n\partial  \\ \end{array} \right).
\label{Kj}
\end{eqnarray}

\begin{lemma}
All the above operators are Hamiltonian operators.
\end{lemma}
{\bf Proof} \quad
It is obvious that $J_1$, $J_2$ and $J_3$ are Hamiltonian operators since they are skew-symmetric operators with constant-coefficient.
It is easy to check $K_1$ and $K_2$ are skew-symmetric. We need to prove that both $K_1$ and $K_2$ satisfy the Jacobi identities 
\begin{eqnarray}
\langle \alpha, K_1'[K_1\beta]\gamma\rangle+cycle (\alpha,\beta,\gamma)=0,
\label{Jacb1}
\\
\langle \alpha, K_2'[K_2\beta]\gamma\rangle+cycle (\alpha,\beta,\gamma)=0,
\label{Jacb2}
\end{eqnarray}
where
$\alpha=(\alpha_1,\alpha_2)^T$, $\beta=(\beta_1,\beta_2)^T$, $\gamma=(\gamma_1,\gamma_2)^T$ are arbitrary vector functions,
the symbol cycle$(\alpha,\beta,\gamma)$ means the cyclic permutation of $\alpha$, $\beta$, $\gamma$,
and the prime-sign means the G\^{a}teaux derivative of an operator $F$ on $q$ in the direction $\sigma$ defined as \cite{Fu}
\begin{equation}
F'[\sigma]=F'(q)[\sigma]=\left.\frac{\partial}{\partial \epsilon}\right|_{\epsilon=0}F(q+\epsilon \sigma).
\label{alpha}
\end{equation}

Let us first prove the Jacobi identity (\ref{Jacb1}). For brevity, we introduce the notations
\begin{eqnarray}
\begin{split}
A=\partial^{-1}(m\alpha_{1}-n\alpha_{2}),\quad B=\partial^{-1}(m\beta_{1}-n\beta_{2}),\quad C=\partial^{-1}(m\gamma_{1}-n\gamma_{2}).
\end{split}
\end{eqnarray}
Direct calculations give rise to
\begin{eqnarray}
\langle \alpha, K_1'[K_1\beta]\gamma\rangle=\int_{-\infty}^{+\infty}[(\alpha_1mB+\alpha_2nB)C+(\alpha_{1}m-\alpha_{2}n)\partial^{-1}(\gamma_1mB+\gamma_2nB)]dx.
\label{Jacb11}
\end{eqnarray}
Integrating (\ref{Jacb11}) by parts, we obtain
\begin{eqnarray}
\begin{split}
\langle \alpha, K_1'[K_1\beta]\gamma\rangle&=\int_{-\infty}^{+\infty}[(\alpha_1mB+\alpha_2nB)C-(\gamma_1mB+\gamma_2nB)\partial^{-1}(\alpha_{1}m-\alpha_{2}n)]dx
\\&=\int_{-\infty}^{+\infty}[(\alpha_1mB+\alpha_2nB)C-(\gamma_1mB+\gamma_2nB)A]dx
\\&=\int_{-\infty}^{+\infty}[(\alpha_1m+\alpha_2n)BC-(\gamma_1m+\gamma_2n)BA]dx.
\end{split}
\label{Jacb12}
\end{eqnarray}
Thus
\begin{eqnarray}
\begin{split}
&\langle \alpha, K_1'[K_1\beta]\gamma\rangle+cycle (\alpha,\beta,\gamma)
\\=&\int_{-\infty}^{+\infty}[(\alpha_1m+\alpha_2n)BC-(\gamma_1m+\gamma_2n)BA]dx
\\&+\int_{-\infty}^{+\infty}[(\beta_1m+\beta_2n)CA-(\alpha_1m+\alpha_2n)CB]dx
\\&+\int_{-\infty}^{+\infty}[(\gamma_1m+\gamma_2n)AB-(\beta_1m+\beta_2n)AC]dx
\\=&0.
\end{split}
\label{Jacb13}
\end{eqnarray}

Now we turn to the proof of Jacobi identity (\ref{Jacb2}). Let us set
\begin{eqnarray}
\begin{split}
\tilde{A}=\partial^{-1}(m\alpha_{1,x}+n\alpha_{2,x}),\quad \tilde{B}=\partial^{-1}(m\beta_{1,x}+n\beta_{2,x}),\quad \tilde{C}=\partial^{-1}(m\gamma_{1,x}+n\gamma_{2,x}).
\end{split}
\end{eqnarray}
With the similar calculations as (\ref{Jacb11}) and (\ref{Jacb12}), we arrive at
\begin{eqnarray}
\langle \alpha, K_2'[K_2\beta]\gamma\rangle=\int_{-\infty}^{+\infty}[(\gamma_{1,x}m_x+\gamma_{2,x}n_x)\tilde{B}\tilde{A}-(\alpha_{1,x}m_x+\alpha_{2,x}n_x)\tilde{B}\tilde{C}
+\tilde{C_x}\tilde{B_x}\tilde{A}-\tilde{A_x}\tilde{B_x}\tilde{C}]dx.
\label{Jacb21}
\end{eqnarray}
Then it follows that
\begin{eqnarray}
\begin{split}
&\langle \alpha, K_2'[K_2\beta]\gamma\rangle+cycle (\alpha,\beta,\gamma)
\\=&\int_{-\infty}^{+\infty}[(\gamma_{1,x}m_x+\gamma_{2,x}n_x)\tilde{B}\tilde{A}-(\alpha_{1,x}m_x+\alpha_{2,x}n_x)\tilde{B}\tilde{C}
+\tilde{C_x}\tilde{B_x}\tilde{A}-\tilde{A_x}\tilde{B_x}\tilde{C}]dx
\\&+\int_{-\infty}^{+\infty}[(\alpha_{1,x}m_x+\alpha_{2,x}n_x)\tilde{C}\tilde{B}-(\beta_{1,x}m_x+\beta_{2,x}n_x)\tilde{C}\tilde{A}
+\tilde{A_x}\tilde{C_x}\tilde{B}-\tilde{B_x}\tilde{C_x}\tilde{A}]dx
\\&+\int_{-\infty}^{+\infty}[(\beta_{1,x}m_x+\beta_{2,x}n_x)\tilde{A}\tilde{C}-(\gamma_{1,x}m_x+\gamma_{2,x}n_x)\tilde{A}\tilde{B}
+\tilde{B_x}\tilde{A_x}\tilde{C}-\tilde{C_x}\tilde{A_x}\tilde{B}]dx
\\=&0.
\end{split}
\label{J22}
\end{eqnarray}
The proof of lemma 1 is finished.

\begin{lemma} The following relations hold
\begin{eqnarray}
\langle \alpha, J_1'[J_2\beta]\gamma\rangle+\langle \alpha, J_2'[J_1\beta]\gamma\rangle+cycle (\alpha,\beta,\gamma)&=&0,
\label{lm21}
\\
\langle \alpha, J_1'[J_3\beta]\gamma\rangle+\langle \alpha, J_3'[J_1\beta]\gamma\rangle+cycle (\alpha,\beta,\gamma)&=&0,
\label{lm22}
\\
\langle \alpha, J_2'[J_3\beta]\gamma\rangle+\langle \alpha, J_3'[J_2\beta]\gamma\rangle+cycle (\alpha,\beta,\gamma)&=&0,
\label{lm23}
\\
\langle \alpha, J_1'[K_1\beta]\gamma\rangle+\langle \alpha, K_1'[J_1\beta]\gamma\rangle+cycle (\alpha,\beta,\gamma)&=&0,
\label{lm24}
\\
\langle \alpha, J_2'[K_1\beta]\gamma\rangle+\langle \alpha, K_1'[J_2\beta]\gamma\rangle+cycle (\alpha,\beta,\gamma)&=&0,
\label{lm25}
\\
\langle \alpha, J_2'[K_2\beta]\gamma\rangle+\langle \alpha, K_2'[J_2\beta]\gamma\rangle+cycle (\alpha,\beta,\gamma)&=&0,
\label{lm26}
\\
\langle \alpha, J_3'[K_2\beta]\gamma\rangle+\langle \alpha, K_2'[J_3\beta]\gamma\rangle+cycle (\alpha,\beta,\gamma)&=&0,
\label{lm27}
\\
\langle \alpha, K_1'[K_2\beta]\gamma\rangle+\langle \alpha, K_2'[K_1\beta]\gamma\rangle+cycle (\alpha,\beta,\gamma)&=&0.
\label{lm28}
\end{eqnarray}
\end{lemma}
{\bf Proof} \quad It is clearly formulas (\ref{lm21})-(\ref{lm23}) hold since $J_1$, $J_2$ and $J_3$ are constant-coefficient operators. For (\ref{lm24}), we have
$\langle \alpha, J_1'[K_1\beta]\gamma\rangle$=0, and
$$\langle \alpha, K_1'[J_1\beta]\gamma\rangle=\int_{-\infty}^{+\infty}[(\alpha_1\beta_2+\alpha_2\beta_1)C-(\beta_1\gamma_2+\beta_2\gamma_1)A]dx.$$
Thus the left hand side (LHS) of (\ref{lm24}) becomes
\begin{eqnarray*}
&&\int_{-\infty}^{+\infty}[(\alpha_1\beta_2+\alpha_2\beta_1)C-(\beta_1\gamma_2+\beta_2\gamma_1)A]dx+\int_{-\infty}^{+\infty}[(\beta_1\gamma_2+\beta_2\gamma_1)A-(\gamma_1\alpha_2+\gamma_2\alpha_1)B]dx
\\&&+
\int_{-\infty}^{+\infty}[(\gamma_1\alpha_2+\gamma_2\alpha_1)B-(\alpha_1\beta_2+\alpha_2\beta_1)C]dx
\\&=&0.
\end{eqnarray*}
Similarly, by direct calculations, the LHS of (\ref{lm25}) becomes
\begin{eqnarray*}
\begin{split}
&\int_{-\infty}^{+\infty}[(\beta_{1,x}\alpha_2-\beta_{2,x}\alpha_1)C-(\beta_{1,x}\gamma_2-\beta_{2,x}\gamma_1)A]dx
\\+&\int_{-\infty}^{+\infty}[(\gamma_{1,x}\beta_2-\gamma_{2,x}\beta_1)A-(\gamma_{1,x}\alpha_2-\gamma_{2,x}\alpha_1)B]dx
\\+&
\int_{-\infty}^{+\infty}[(\alpha_{1,x}\gamma_2-\alpha_{2,x}\gamma_1)B-(\alpha_{1,x}\beta_2-\alpha_{2,x}\beta_1)C]dx
\\=&\int_{-\infty}^{+\infty}[(\beta_{1}\alpha_2-\beta_{2}\alpha_1)_xC-(\beta_{1}\gamma_2-\beta_{2}\gamma_1)_xA-(\gamma_{1}\alpha_2-\gamma_{2}\alpha_1)_xB]dx
\\=&-\int_{-\infty}^{+\infty}[(\beta_{1}\alpha_2-\beta_{2}\alpha_1)C_x-(\beta_{1}\gamma_2-\beta_{2}\gamma_1)A_x-(\gamma_{1}\alpha_2-\gamma_{2}\alpha_1)B_x]dx
\\=&-\int_{-\infty}^{+\infty}[(\beta_{1}\alpha_2-\beta_{2}\alpha_1)(m\gamma_{1}-n\gamma_{2})-(\beta_{1}\gamma_2-\beta_{2}\gamma_1)(m\alpha_{1}-n\alpha_{2})-(\gamma_{1}\alpha_2-\gamma_{2}\alpha_1)(m\beta_{1}-n\beta_{2})]dx
\\=&0.
\end{split}
\end{eqnarray*}
The LHS of (\ref{lm26}) is equal to
\begin{eqnarray*}
\begin{split}
&-\int_{-\infty}^{+\infty}[(\alpha_{1,x}\beta_{2,x}+\alpha_{2,x}\beta_{1,x})\tilde{C}-(\beta_{2,x}\gamma_{1,x}+\beta_{1,x}\gamma_{2,x})\tilde{A}]dx
\\&-\int_{-\infty}^{+\infty}[(\beta_{1,x}\gamma_{2,x}+\beta_{2,x}\gamma_{1,x})\tilde{A}-(\gamma_{2,x}\alpha_{1,x}+\gamma_{1,x}\alpha_{2,x})\tilde{B}]dx
\\&-\int_{-\infty}^{+\infty}[(\gamma_{1,x}\alpha_{2,x}+\gamma_{2,x}\alpha_{1,x})\tilde{B}-(\alpha_{2,x}\beta_{1,x}+\alpha_{1,x}\beta_{2,x})\tilde{C}]dx
\\=&0.
\end{split}
\end{eqnarray*}
The LHS of (\ref{lm27}) is equal to
\begin{eqnarray*}
\begin{split}
&-\int_{-\infty}^{+\infty}[(\alpha_{1,x}\beta_{2,xx}-\alpha_{2,x}\beta_{1,xx})\tilde{C}-(\beta_{2,xx}\gamma_{1,x}-\beta_{1,xx}\gamma_{2,x})\tilde{A}]dx
\\&-\int_{-\infty}^{+\infty}[(\beta_{1,x}\gamma_{2,xx}-\beta_{2,x}\gamma_{1,xx})\tilde{A}-(\gamma_{2,xx}\alpha_{1,x}-\gamma_{1,xx}\alpha_{2,x})\tilde{B}]dx
\\&-\int_{-\infty}^{+\infty}[(\gamma_{1,x}\alpha_{2,xx}-\gamma_{2,x}\alpha_{1,xx})\tilde{B}-(\alpha_{2,xx}\beta_{1,x}-\alpha_{1,xx}\beta_{2,x})\tilde{C}]dx
\\=&-\int_{-\infty}^{+\infty}[(\beta_{1,x}\gamma_{2,x}-\beta_{2,x}\gamma_{1,x})_{x}\tilde{A}+(\gamma_{1,x}\alpha_{2,x}-\gamma_{2,x}\alpha_{1,x})_x\tilde{B}+(\alpha_{1,x}\beta_{2,x}-\alpha_{2,x}\beta_{1,x})_x\tilde{C}]dx
\\=&\int_{-\infty}^{+\infty}[(\beta_{1,x}\gamma_{2,x}-\beta_{2,x}\gamma_{1,x})\tilde{A}_{x}+(\gamma_{1,x}\alpha_{2,x}-\gamma_{2,x}\alpha_{1,x})\tilde{B}_{x}+(\alpha_{1,x}\beta_{2,x}-\alpha_{2,x}\beta_{1,x})\tilde{C}_{x}]dx
\\=&\int_{-\infty}^{+\infty}[(\beta_{1,x}\gamma_{2,x}-\beta_{2,x}\gamma_{1,x})(m\alpha_{1,x}+n\alpha_{2,x})+(\gamma_{1,x}\alpha_{2,x}-\gamma_{2,x}\alpha_{1,x})(m\beta_{1,x}+n\beta_{2,x})
\\&~~~~~~~~+(\alpha_{1,x}\beta_{2,x}-\alpha_{2,x}\beta_{1,x})(m\gamma_{1,x}+n\gamma_{2,x})]dx
\\=&0.
\end{split}
\end{eqnarray*}
For the LHS of (\ref{lm28}), we have
\begin{eqnarray*}\langle \alpha, K_1'[K_2\beta]\gamma\rangle&=&\int_{-\infty}^{+\infty}[(m\alpha_{1,x}-n\alpha_{2,x})C\tilde{B}-(m\gamma_{1,x}-n\gamma_{2,x})A\tilde{B}]dx,
\\
\langle \alpha, K_2'[K_1\beta]\gamma\rangle&=&\int_{-\infty}^{+\infty}[(m\alpha_{1,x}-n\alpha_{2,x})B\tilde{C}-(m\gamma_{1,x}-n\gamma_{2,x})B\tilde{A}]dx.
\end{eqnarray*}
Hence the LHS of (\ref{lm28}) reads as
\begin{eqnarray*}
\begin{split}
&\int_{-\infty}^{+\infty}[(m\alpha_{1,x}-n\alpha_{2,x})(C\tilde{B}+B\tilde{C})-(m\gamma_{1,x}-n\gamma_{2,x})(A\tilde{B}+B\tilde{A})]dx
\\+&\int_{-\infty}^{+\infty}[(m\beta_{1,x}-n\beta_{2,x})(A\tilde{C}+C\tilde{A})-(m\alpha_{1,x}-n\alpha_{2,x})(B\tilde{C}+C\tilde{B})]dx
\\+&\int_{-\infty}^{+\infty}[(m\gamma_{1,x}-n\gamma_{2,x})(B\tilde{A}+A\tilde{B})-(m\beta_{1,x}-n\beta_{2,x})(C\tilde{A}+A\tilde{C})]dx
\\=&0.
\end{split}
\end{eqnarray*}
This completes the proof of lemma 2.

Lemma 2 implies that $J_1+J_2$, $J_1+J_3$, $J_1+K_1$, $J_2+J_3$, $J_2+K_1$, $J_2+K_2$, $J_3+K_2$ and $K_1+K_2$ are Hamiltonian operators. But we should notice that $J_1+K_2$ and $J_3+K_1$ are not Hamiltonian operators. In fact, we have
\begin{lemma} The following two relations hold
\begin{eqnarray}
\begin{split}
&\langle \alpha, J_1'[K_2\beta]\gamma\rangle+\langle \alpha, K_2'[J_1\beta]\gamma\rangle+cycle (\alpha,\beta,\gamma)
\\
=&\int_{-\infty}^{+\infty}[(\beta_1\gamma_2-\beta_2\gamma_1)_x\tilde{A}+(\alpha_2\gamma_1-\alpha_1\gamma_2)_x\tilde{B}+(\alpha_1\beta_2-\alpha_2\beta_1)_x\tilde{C}]dx,
\\
&\langle \alpha, J_3'[K_1\beta]\gamma\rangle+\langle \alpha, K_1'[J_3\beta]\gamma\rangle+cycle (\alpha,\beta,\gamma)
\\
=&\int_{-\infty}^{+\infty}[(\beta_{1,x}\gamma_2-\beta_1\gamma_{2,x}+\beta_{2,x}\gamma_1-\beta_2\gamma_{1,x})_xA+(\gamma_{1,x}\alpha_2-\gamma_1\alpha_{2,x}+\gamma_{2,x}\alpha_1-\gamma_2\alpha_{1,x})_xB
\\&~~~~~~~~+(\alpha_{1,x}\beta_2-\alpha_1\beta_{2,x}+\alpha_{2,x}\beta_1-\alpha_2\beta_{1,x})_xC]dx.
\label{lm3}
\end{split}
\end{eqnarray}
\end{lemma}
{\bf Proof} \quad Direct calculations yield that
\begin{eqnarray*}
\begin{split}
&\langle \alpha, J_1'[K_2\beta]\gamma\rangle+\langle \alpha, K_2'[J_1\beta]\gamma\rangle+cycle (\alpha,\beta,\gamma)
\\
=&-\int_{-\infty}^{+\infty}[(\beta_{1}\alpha_{2,x}-\beta_{2}\alpha_{1,x})\tilde{C}-(\beta_{1}\gamma_{2,x}-\beta_{2}\gamma_{1,x})\tilde{A}]dx
\\&-\int_{-\infty}^{+\infty}[(\gamma_{1}\beta_{2,x}-\gamma_{2}\beta_{1,x})\tilde{A}-(\gamma_{1}\alpha_{2,x}-\gamma_{2}\alpha_{1,x})\tilde{B}]dx
\\&-\int_{-\infty}^{+\infty}[(\alpha_{1}\gamma_{2,x}-\alpha_{2}\gamma_{1,x})\tilde{B}-(\alpha_{1}\beta_{2,x}-\alpha_{2}\beta_{1,x})\tilde{C}]dx
\\
=&\int_{-\infty}^{+\infty}[(\beta_1\gamma_2-\beta_2\gamma_1)_x\tilde{A}+(\alpha_2\gamma_1-\alpha_1\gamma_2)_x\tilde{B}+(\alpha_1\beta_2-\alpha_2\beta_1)_x\tilde{C}]dx,
\end{split}
\end{eqnarray*}
and
\begin{eqnarray*}
\begin{split}
&\langle \alpha, J_3'[K_1\beta]\gamma\rangle+\langle \alpha, K_1'[J_3\beta]\gamma\rangle+cycle (\alpha,\beta,\gamma)
\\
=&-\int_{-\infty}^{+\infty}[(\alpha_{1}\beta_{2,xx}+\alpha_{2}\beta_{1,xx})C-(\beta_{1,xx}\gamma_2+\beta_{2,xx}\gamma_1)A]dx
\\
&-\int_{-\infty}^{+\infty}[(\beta_{1}\gamma_{2,xx}+\beta_{2}\gamma_{1,xx})A-(\gamma_{1,xx}\alpha_2+\gamma_{2,xx}\alpha_1)B]dx
\\
&-\int_{-\infty}^{+\infty}[(\gamma_{1}\alpha_{2,xx}+\gamma_{2}\alpha_{1,xx})B-(\alpha_{1,xx}\beta_2+\alpha_{2,xx}\beta_1)C]dx
\\
=&\int_{-\infty}^{+\infty}[(\beta_{1,x}\gamma_2-\beta_1\gamma_{2,x}+\beta_{2,x}\gamma_1-\beta_2\gamma_{1,x})_xA+(\gamma_{1,x}\alpha_2-\gamma_1\alpha_{2,x}+\gamma_{2,x}\alpha_1-\gamma_2\alpha_{1,x})_xB
\\&~~~~~~~~+(\alpha_{1,x}\beta_2-\alpha_1\beta_{2,x}+\alpha_{2,x}\beta_1-\alpha_2\beta_{1,x})_xC]dx.
\end{split}
\end{eqnarray*}
This finishes the proof of lemma 3.

\begin{lemma} The following Jacobi identity holds
\begin{eqnarray}
\begin{split}
\langle \alpha, J_1'[K_2\beta]\gamma\rangle+\langle \alpha, K_2'[J_1\beta]\gamma\rangle+\langle \alpha, J_3'[K_1\beta]\gamma\rangle+\langle \alpha, K_1'[J_3\beta]\gamma\rangle+cycle (\alpha,\beta,\gamma)
=0.
\label{lm4}
\end{split}
\end{eqnarray}
\end{lemma}
{\bf Proof} \quad
By virtue of lemma 3 and integration by parts, we arrive at
\begin{eqnarray*}
\begin{split}
&\langle \alpha, J_1'[K_2\beta]\gamma\rangle+\langle \alpha, K_2'[J_1\beta]\gamma\rangle+\langle \alpha, J_3'[K_1\beta]\gamma\rangle+\langle \alpha, K_1'[J_3\beta]\gamma\rangle+cycle (\alpha,\beta,\gamma)
\\=&
\int_{-\infty}^{+\infty}[(\beta_1\gamma_2-\beta_2\gamma_1)_x\tilde{A}+(\alpha_2\gamma_1-\alpha_1\gamma_2)_x\tilde{B}+(\alpha_1\beta_2-\alpha_2\beta_1)_x\tilde{C}
\\
&~~~~~~~~+(\beta_{1,x}\gamma_2-\beta_1\gamma_{2,x}+\beta_{2,x}\gamma_1-\beta_2\gamma_{1,x})_xA+(\gamma_{1,x}\alpha_2-\gamma_1\alpha_{2,x}+\gamma_{2,x}\alpha_1-\gamma_2\alpha_{1,x})_xB
\\&~~~~~~~~+(\alpha_{1,x}\beta_2-\alpha_1\beta_{2,x}+\alpha_{2,x}\beta_1-\alpha_2\beta_{1,x})_xC]dx
\\=&
-\int_{-\infty}^{+\infty}[(\beta_1\gamma_2-\beta_2\gamma_1)\tilde{A}_x+(\alpha_2\gamma_1-\alpha_1\gamma_2)\tilde{B}_x+(\alpha_1\beta_2-\alpha_2\beta_1)\tilde{C}_x
\\
&~~~~~~~~~~~~+(\beta_{1,x}\gamma_2-\beta_1\gamma_{2,x}+\beta_{2,x}\gamma_1-\beta_2\gamma_{1,x})A_x+(\gamma_{1,x}\alpha_2-\gamma_1\alpha_{2,x}+\gamma_{2,x}\alpha_1-\gamma_2\alpha_{1,x})B_x
\\&~~~~~~~~~~~~+(\alpha_{1,x}\beta_2-\alpha_1\beta_{2,x}+\alpha_{2,x}\beta_1-\alpha_2\beta_{1,x})C_x]dx
\\=&
-\int_{-\infty}^{+\infty}[(\beta_1\gamma_2-\beta_2\gamma_1)(m\alpha_{1,x}+n\alpha_{2,x})+(\alpha_2\gamma_1-\alpha_1\gamma_2)(m\beta_{1,x}+n\beta_{2,x})
\\&~~~~~~~~~~~~+(\alpha_1\beta_2-\alpha_2\beta_1)(m\gamma_{1,x}+n\gamma_{2,x})+(\beta_{1,x}\gamma_2-\beta_1\gamma_{2,x}+\beta_{2,x}\gamma_1-\beta_2\gamma_{1,x})(m\alpha_{1}-n\alpha_{2})
\\&~~~~~~~~~~~~+(\gamma_{1,x}\alpha_2-\gamma_1\alpha_{2,x}+\gamma_{2,x}\alpha_1-\gamma_2\alpha_{1,x})(m\beta_{1}-n\beta_{2})
\\&~~~~~~~~~~~~+(\alpha_{1,x}\beta_2-\alpha_1\beta_{2,x}+\alpha_{2,x}\beta_1-\alpha_2\beta_{1,x})(m\gamma_{1}-n\gamma_{2})]dx
\\
=&0.
\label{lm41}
\end{split}
\end{eqnarray*}
 The proof of lemma 4 is finished.

Based on the above lemmas, we finally obtain
\begin{proposition}
Let $c_j$, $1\leq j\leq 5$, be arbitrary constants. For any $c_1c_5=c_3c_4$, we can conclude that $J=c_1J_1+c_2J_2+c_3J_3+c_4K_1+c_5K_2$ is a Hamiltonian operator.
\end{proposition}
{\bf Proof} \quad
We need to verify the Jacobi identity
\begin{eqnarray*}
\langle \alpha, J'[J\beta]\gamma\rangle+cycle (\alpha,\beta,\gamma)=0.
\end{eqnarray*}
In fact, we have
\begin{eqnarray*}
\begin{split}
&\langle \alpha, J'[J\beta]\gamma\rangle+cycle (\alpha,\beta,\gamma)
\\=&c_1^2\langle \alpha, J_1'[J_1\beta]\gamma\rangle
+c_2^2\langle \alpha, J_2'[J_2\beta]\gamma\rangle
+c_3^2\langle \alpha, J_3'[J_3\beta]\gamma\rangle
+c_4^2\langle \alpha, K_1'[K_1\beta]\gamma\rangle
+c_5^2\langle \alpha, K_2'[K_2\beta]\gamma\rangle
\\&+c_1c_2\left(\langle \alpha, J_1'[J_2\beta]\gamma\rangle+\langle \alpha, J_2'[J_1\beta]\gamma\rangle\right)
+c_1c_3\left(\langle \alpha, J_1'[J_3\beta]\gamma\rangle+\langle \alpha, J_3'[J_1\beta]\gamma\rangle\right)
\\&+c_1c_4\left(\langle \alpha, J_1'[K_1\beta]\gamma\rangle+\langle \alpha, K_1'[J_1\beta]\gamma\rangle\right)+c_1c_5\left(\langle \alpha, J_1'[K_2\beta]\gamma\rangle+\langle \alpha, K_2'[J_1\beta]\gamma\rangle\right)
\\&+c_2c_3\left(\langle \alpha, J_2'[J_3\beta]\gamma\rangle+\langle \alpha, J_3'[J_2\beta]\gamma\rangle\right)
+c_2c_4\left(\langle \alpha, J_2'[K_1\beta]\gamma\rangle+\langle \alpha, K_1'[J_2\beta]\gamma\rangle\right)
\\&+c_2c_5\left(\langle \alpha, J_2'[K_2\beta]\gamma\rangle+\langle \alpha, K_2'[J_2\beta]\gamma\rangle\right)
+c_3c_4\left(\langle \alpha, J_3'[K_1\beta]\gamma\rangle+\langle \alpha, K_1'[J_3\beta]\gamma\rangle\right)
\\&+c_3c_5\left(\langle \alpha, J_3'[K_2\beta]\gamma\rangle+\langle \alpha, K_2'[J_3\beta]\gamma\rangle\right)
+c_4c_5\left(\langle \alpha, K_1'[K_2\beta]\gamma\rangle+\langle \alpha, K_2'[K_1\beta]\gamma\rangle\right)
\\&+cycle (\alpha,\beta,\gamma)
\\=&c_1c_5\left(\langle \alpha, J_1'[K_2\beta]\gamma\rangle+\langle \alpha, K_2'[J_1\beta]\gamma\rangle\right)
+c_3c_4\left(\langle \alpha, J_3'[K_1\beta]\gamma\rangle+\langle \alpha, K_1'[J_3\beta]\gamma\rangle\right)
+cycle (\alpha,\beta,\gamma)
\\=&0,
\end{split}
\end{eqnarray*}
where the last identity holds because of $c_1c_5=c_3c_4$ and lemma 4. This completes the proof of the proposition.

Recall that a pair of Hamiltonian operators $J$ and $K$ is called compatible, if $J+K$ is Hamiltonian.
From the above proposition, we immediately arrive at
\begin{corollary}
The case of $c_1=-c_2=c_4=1$ and $c_3=c_5=0$ leads to the compatibility of the Hamiltonian operators (\ref{JK0}).
\end{corollary}
\begin{corollary}
The case of $c_1=c_3=c_4=c_5=1$ and $c_2=0$ leads to the compatibility of the Hamiltonian operators (\ref{JK1}).
\end{corollary}
\begin{corollary}
The case of $c_2=c_3=c_5=1$ and $c_1=c_4=0$ leads to the compatibility of the Hamiltonian operators (\ref{JK2}).
\end{corollary}

\vspace*{0.2cm}

{\bf Remark 6.} The Hamiltonian pair $J$ and $K$ in each example in section 3 is a special case of the generalized form $J=c_1J_1+c_2J_2+c_3J_3$ and $K=c_4K_1+c_5K_2$, where $c_1c_5=c_3c_4$.
The compatibility of such a Hamiltonian pair is guaranteed by proposition 1.

\section {Conclusions and discussions}
In the paper, from the spectral problems (\ref{lps}) and (\ref{lpt}), we propose a generalized two-component model (\ref{geq}) which allows for an arbitrary function $H$ to be involved in.
We may generate many integrable peakon systems with different choices of $H$ in our model. So, our model provides a large class of peakon systems and covers almost all existing integrable peakon equations associated with $2\times 2$ spectral problems. Because of the presence of an arbitrary function in the generalized system, we do not expect all those equations possess the bi-Hamiltonian structures in general. Nevertheless, we show that for some special choices of the function $H$ in (\ref{geq}) we may find the bi-Hamiltonian structures. Moreover, from the generalized model we obtain very interesting solutions, such as new type of $N$-peakon solution which is not in the traveling wave type.

Different from the usual integrable soliton equations, the peakon equation involved in an arbitrary function seems to be unusual. We believe that this system deserves a further investigation.
The following two problems seem to be interesting:

$\bullet$ Is there a gauge transformation that can remove the arbitrary function $H$?

$\bullet$ Can the inverse scattering transforms be applied to solve our system in general?


Very recently, we know that Li, Liu and Popowicz \cite{LLP} proposed a four-component peakon equation with an arbitrary function involved in,
where they cited a preprint version \cite{XQZ} of the present paper. We believe that both our generalized peakon system and Li-Liu-Popowicz's system deserve a further investigation.


\section*{ACKNOWLEDGMENTS}
The authors would like to express their sincerest thanks to the anonymous referee for the helpful suggestions and invaluable comments, which have helped us to improve this paper.
The authors Xia and Zhou were supported by the National Natural Science Foundation of China (Grant Nos. 11301229 and 11271168), the Natural Science Foundation of the Jiangsu Province (Grant No. BK20130224) and the Natural Science Foundation of the Jiangsu Higher Education Institutions of China (Grant No. 13KJB110009). The author Qiao was partially supported by the National Natural Science Foundation of China (No. 11171295, No. 61301187, and No. 61328103) and also thanks the U.S. Department of Education GAANN project (P200A120256) to support UTPA mathematics graduate program.

\vspace{1cm}
\small{

}
\end{document}